# A Generalized Signal Quality Estimation Method for IoT Sensors

Arlene John[1], Barry Cardiff[1,2], *Senior Member, IEEE*, and Deepu John[1,2], *Senior Member, IEEE*
[1]University College Dublin, Ireland , [2]Microelectronic Circuits Centre Ireland
Email: arlene.john@ucdconnect.ie, barry.cardiff@ucd.ie, deepu.john@ucd.ie

*Abstract*— IoT wearable devices are widely expected to reduce the cost and risk of personal healthcare. However, ambulatory data collected from such devices are often corrupted or contaminated with severe noises. Signal Quality Indicators (SQIs) can be used to assess the quality of data obtained from wearable devices, such that transmission/ storage of unusable data can be prevented. This article introduces a novel and generalized SQI which can be implemented on an edge device for detecting the quality of any quasi-periodic signal under observation, regardless of the type of noise present. The application of this SQI on Electrocardiogram (ECG) signals is investigated. From the analysis carried out, it was found that the proposed generalized SQI is suitable for quality assessment of ECG signals and exhibits a linear behavior in the medium to high SNR regions under all noise conditions considered. The proposed SQI was used for acceptability testing of ECG records in CinC Physionet 2011 challenge dataset and found to be accurate for 90.4% of the records while having minimal computational complexity.

## I. Introduction

The unsustainability of the present-day healthcare delivery model, the prevalence of diseases associated with current sedentary lifestyles and improvements in electronic and signal processing technologies have led to the introduction of wearable devices for continuous health and fitness monitoring [1]. Wearable devices are often prone to motion artifacts and other common environmental noises. The acquired data is used for evaluation of fitness activity or health condition of the user or in specific cases, clinicians make use of this data for detailed clinical analysis. However, the quality of the acquired signal is an important factor in the quality of the signal analysis, especially in the case of wearable physiological signal monitoring devices that give immediate feedback to the user. Therefore, automatic estimation of noisy or clean segments (signal quality) of continuously monitored data at the edge is essential in health and fitness monitoring devices. Recent studies indicate the importance of incorporating SQI in wearable devices for effective physiological activity monitoring [1], with processing data at the edge to 1) reduce data redundancies 2) avoid transmission of corrupted or unusable data 3) optimize storage and battery resources in IoT devices. Methods for indicating the signal quality of few biological signals such as ECG signals [2]–[5], Photoplethsymogram (PPG) signals [1], [6], [7], Arterial Blood pressure signals [8], etc. have been discussed in literature. The 2011 CinC challenge focused on improving the quality of ECG signals obtained via mobile phones aimed to develop techniques to grade ECG signals as of acceptable/ unacceptable quality [9]. Recent literature on data fusion stresses the importance of choosing only clean segments of the signal for fusion [10]. Signal quality indicators can be used as a confidence metric for features extracted and also reduce the cases of false alarms as they stipulate whether the signal can be used for further analysis [11]. A major concern associated with signal quality indicators is that there is no general definition of signal quality since quality requirements are heavily dependent on the application and the further steps of processing involved. For example, if the processing device is not capable of removing power line interference from a particular signal, then a signal segment corrupted with power line interference should be assigned an SQI value corresponding to poor quality. On the other hand, in a more sophisticated signal processing system- where power line interference can be eliminated- even if the signal is corrupted with power line interference but clean in other regards, a high signal quality value could be assigned to that signal segment.

From literature, one can observe that signal quality indicators for a particular kind of signal or noise scenario is usually discussed, increasing the complexity in heterogeneous signal systems. To detect the quality of a signal, the algorithm must be capable of detecting the amount of noise contained in the signal, therefore a good signal quality indicator must be a monotonically increasing function of the signal to noise ratio (SNR), which can then be mapped as a linear function of SNR. But commonly used SQIs, especially for ECG signal quality assessment do not exhibit a monotonically increasing behavior with an increase in SNR.

## II. Curve SQI (cSQI)

In this article, we propose a generalized signal quality indicator that could be made suitable for detecting the signal quality of any periodic or quasi-periodic signals with special application to wearable devices. The SQI proposed is based on the waveform morphology of the signal of interest and tries to evaluate how similar the signal is to the expected waveform shape. This SQI is then tested out on ECG signals and as there is no commonly accepted metric for checking the performance of a signal quality indicator, except the nature of the SQI as a function of SNR, this article would refrain from comparing with other commonly used ECG signal quality indicators as it is beyond the scope of this paper.

### A. Template Generation

The first step in carrying out signal quality analysis using cSQI is the generation of a single cycle template. This is a pre-requisite and must be done in a wearable device during the initialization stage where the user can be prompted to stay still and clear from noise sources. Therefore, it can be assumed that the signal obtained in the initialization stage is clean. The template generation would attempt to use **N** cycles of clean data and the time taken for generation of the template will vary depending on the average time period of a given quasi-periodic signal. (For Eg: In case of ECG signals indicating heart rate, the knowledge that the average heart rate is 60 to 100 beats per minute. Let's consider it to be 80 bpm, therefore 10 cycles can be obtained in approximately 7.5 seconds). The knowledge of this time period also helps us in fixing the template length that needs to be used for template generation. The steps involved in generating a template of length $2M+1$ samples are as in Table 1. Once the signal for template generation is acquired, the template can be generated offline off the device to save power.

**Table 1: Algorithm for Template Generation**

**Setup:**
A. Signal sequence of interest $\mathbf{x}$
B. Signal vector $\mathbf{X_n}$ of length $2M+1 \triangleq [\mathbf{x}[n-M],…\mathbf{x}[n],…\mathbf{x}[n+M]]$ where M has been decided based on the time period.
C. Let $\mathbf{X_{n\,(c)}}$ be the correlation Toeplitz matrix of $\mathbf{X_n}$ of dimensions (2M+1, 4M+1) where $\mathbf{X_{n\,(c)}}=$

$$\begin{bmatrix} \mathbf{x}[n+M] & …\mathbf{x}[n] & …\mathbf{x}[n-M] & …0 & 0 \\ 0 & \mathbf{x}[n+M] & …\mathbf{x}[n] & \mathbf{x}[n-M] & 0 \\ . & . & . & . & . \\ . & . & . & . & . \\ 0 & 0 & ..\mathbf{x}[n+M] & …\mathbf{x}[n] & …\mathbf{x}[n-M] \end{bmatrix}$$

, where the $k^{th}$ column of $\mathbf{X_{n\,(c)}}$ is indicated by $\mathbf{X_n}^{(k)}_{(c)}$ where k ∈ [0, 4M]
D. Let $\mathbf{X_n}^{(t)}$ be $t^{th}$ circular shifted version of $\mathbf{X_n}$ ie., $\mathbf{X_n}^{(t)} = [\mathbf{x}[n-M+t]\ \mathbf{x}[n-M+t+1]…\mathbf{x}[n+M]\ \mathbf{x}[n-M]…\mathbf{x}[n-M+t-1]]$

**Template Generation:**
1. Identify a signal segment containing N consecutive fiducial features of interest (From the time period or otherwise).
2. Let $\mathbf{n_0}$ be the center of the first segment of length $2M+1$, $\mathbf{Xn_0}$.
3. Initialize $\mathbf{T}= \mathbf{Xn_0}$.
4. For $\mathbf{i}=(M+1)$ to $(\mathbf{N}(2M+1)-M)$ in steps of $2M+1$
   a. Let $\mathbf{X_{n_0+i\,(c)}}$ be the correlation Toeplitz matrix of $\mathbf{X_{n_0+i}}$
   b. Let $\mathbf{k} = \text{argmax}_j [|\mathbf{T}.\mathbf{X}^{(j)}_{\mathbf{n_0+i\,(c)}}|]$
   c. If $\mathbf{k} \approx \mathbf{n_0}$
      $\mathbf{T}=\mathbf{T} + \mathbf{X}^{(k)}_{\mathbf{n_0+i}}$
      End
   End
5. Normalize $\mathbf{T}$ based on number of times condition 4.c was satisfied.

At this stage, it is important to limit **N** to small numbers because if many cycles are used in obtaining a template, too much averaging out can smoothen the signal template much more than desired.

*B. cSQI calculation*

Once the template is generated, cSQI can be estimated. cSQI is a signal quality indicator that assigns a quality value for each signal sample. This is assigned based on the quality of the signal window centered at that point. cSQI at a point $t_0$ is defined as the inverse of the variance of the difference between the template and signal window centered at $t_0$ at the point of maximum correlation (the point of maximum correlation is found by circular rotation of the template and then correlation calculation). The SQI at the next point $t_1$ is calculated in a similar manner by shifting the window about the centre to the next sample once. This yields a signal quality level at each sample point, which can be used for weighting in multi-channel data fusion applications. The steps in calculating the cSQI per sample are as in Table 2.

**Table 2: Algorithm for cSQI calculation**

**Initialize and calculate c[0]:**
A. Let $\mathbf{T_{(c)}}$ be the Toeplitz correlation matrix of the template $\mathbf{T}$. $\mathbf{T_{(c)}}$ [u] indicates $u^{th}$ row of the matrix with u ∈ [0, 2M]
B. Signal vector $\mathbf{S_0}$ of length $2M+1 \triangleq [\mathbf{s}[-M],… \mathbf{s}[0],…\mathbf{s}[M]]$ where M has been decided based on the time period, centered at $\mathbf{s}[0]$.
C. Lag sequence $\mathbf{L}=[-M\ -M+1\ …0…M-1\ M]$
D. Calculate $\mathbf{A}'= \mathbf{S_0}\ \mathbf{X}\ \mathbf{T}$ where $\mathbf{A}'$ is the L-R flipped result of the correlation sequence
E. Lag $t=\mathbf{L}[\text{argmax}(\mathbf{A})]$
F. $\mathbf{c}\,[0]= (\text{var}(\mathbf{S_0}-\mathbf{T^{(t)}}))^{-1}$ and $\mathbf{b}=\mathbf{s}[-M]$

**Loop till no signal is read:**
For $\mathbf{i}=1$ to $\mathbf{Z}$ in steps of 1
   a. Window centered at $\mathbf{s}[i]$, $\mathbf{S_i}$ of length 2M+1.
   b. $\mathbf{NR}=\mathbf{s}[i+M].\ \mathbf{T_{(c)}}\,[2M]$ and $\mathbf{FR}=\mathbf{b}.\ \mathbf{T_{(c)}}\,[0]$
   c. Calculate L-R flipped correlation sequence $\mathbf{A}'=(\mathbf{A}'-\mathbf{FR})^{(1)}+\mathbf{NR}$
   d. Lag $t=\mathbf{L}[\text{argmax}(\mathbf{A})]$
   e. $\mathbf{c}\,[i] = (\text{var}(\mathbf{S_k}-\mathbf{T^{(t)}}))^{-1}$ and $\mathbf{b}=\mathbf{s}[i-M]$
End

cSQI= LPF(c), where the low pass filter could be a moving average filter to avoid rapid variations that can be attributed to edge effects of windowing. The block diagram of the algorithm for a circuit implementation is shown in Fig 1.

As discussed in Section I, a good signal quality indicator must be a monotonically increasing function of SNR. The suitability of cSQI for ECG signals as a proof of application is discussed in the next section.

### III. cSQI FOR ECG SIGNALS

ECG signals can be corrupted by a variety of noises. The most common examples of noise corrupting ECG signals are Motion Artifacts, Electrode Motion Noise, Muscle Artifacts, Baseline Wander, Power line Interference and Instrumentation noise due to electronic components in the acquisition system. Signal corruption in ECG signals due to leads that have fallen off etc. can be modeled as the ECG signal being corrupted by an additive noise signal. The cSQI can be used to determine which portions of the signals are suitable for further processing (like QRS detection [12] or beat classification [13] in ECG signals) on-chip due to its low complexity as discussed in Section III. C [14] [15].

ECG signal data for checking the suitability of cSQI for estimating ECG signal quality was taken from the MIMIC III waveform database [16], [17]. The experiment was tested on one clean ECG record and the average over 10 noise instances was used to evaluate the performance of the SQIs at different signal to noise ratios under different noise conditions. The different noise conditions considered in this analysis are Electrode Motion noise, Muscle Artifacts, Baseline Wander (for which the recordings were taken from the Noise Stress Test Database [17], [18]), 50 Hz power line interference (modeled as sinusoids and a combination of sinusoids) and white Gaussian noise.

*A. cSQI template generation for ECG signals.*

For generating a template for ECG signals, a clean record from the MIMIC III database was used. The first 10 cycles of the record were used for generating a template. The template length $T$ was chosen as 70% of the average time period as we didn't want to risk the chance of having 2 fiducial points in one segment of data in case the time period decreases.
In scenarios, when the segment completely misses a fiducial point, that segment can then be discarded due to condition 4.c in Table 1. Because of this condition, all 10 cycles may not contribute to the final template. The MIMIC III database records signals at a sampling rate of 125 Hz and from the time, the chosen template was of length 71, with M=35. The samples from the $0^{th}$ to the $674^{th}$ sample were used for template generation while the cSQI performance evaluation was carried out on the rest of the signal. For the ECG signals, the fiducial point was chosen at the Q wave which is the maxima in one cycle of an ECG signal. Fig 2 shows the overlay of ECG cycles obtained from the first 10 cycles that are averaged to generate the template.

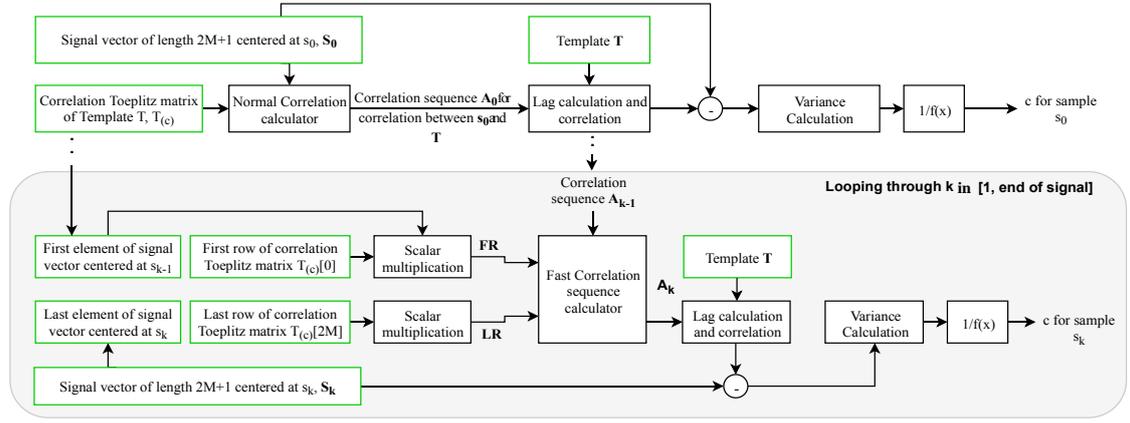

Fig. 1. Block diagram for implementing cSQI calculation (instantaneous c values). The boxes with green outline show the inputs.

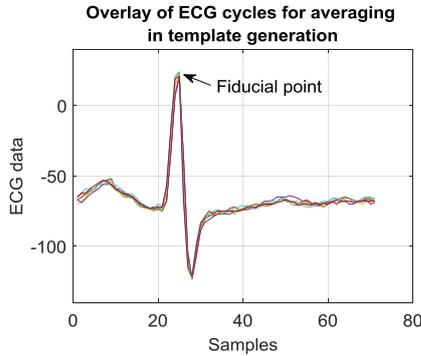

Fig. 2. Overlay of selected cycles for template generation

### B. Calculation of cSQI for an ECG signal

cSQI at each sample point is calculated based on the algorithm in Table 2 in Section II.B. In Fig 3, the intricacies of how a window is selected and how the template curve is circularly rotated to match with the points of maximum correlation in the signal window is illustrated. Fig 3 also shows how the cSQI calculation varies when the signal is corrupted by 5 dB Gaussian noise and when the signal is clean and the difference in values of the SQIs is also indicated. From Fig 3, we can observe how the template generated in Section III. A is circularly rotated to align with the fiducial feature in the signal window such that the difference between the corrected template and the signal window is minimal. Therefore, the difference between the signal window and the corrected template would correspond to noise (Observe in Fig 3, when the signal window is noisy and the template is overlaid on it, the difference between these two signal vectors is the added noise).

### C. Performance Evaluation

For clarity of understanding, the cSQI plots at each sample point for when a portion of the signal is corrupted by (a) 50 Hz power line interference, (b) Baseline Wander, (c) Muscle Artifacts and (d) Electrode Motion at -10 dB is shown in Fig 4. Observe in the figure how cSQI values drop when the signal is noisy compared to when the signal is clean. From this figure, we can observe that at for all samples of the noisy signal, the cSQI calculated is much lower than that of when the signal is clean, which is a simple but good indicator of a good SQI. The cSQI vs SNR plot for when the ECG signal is corrupted by different noise conditions is shown in Fig 5.

Here the SQI is averaged over the entire signal per noise instance per SNR. Most ideally, the SQI values must vary linearly with SNR, but since this is not possible as signal quality can vary from 0 to ∞, a monotonically increasing function is a sufficiently good estimate of the signal quality. But it can be observed from Fig 5 that the SQI exhibits an almost linear behavior in the medium SNR ranges before saturating completely at the high SNR ranges. From the figure, we can conclude that curve based SQI is a sufficiently good indicator of signal quality as the function is monotonically increasing against the SNR values and the behavior is close to identical for all noise conditions considered.

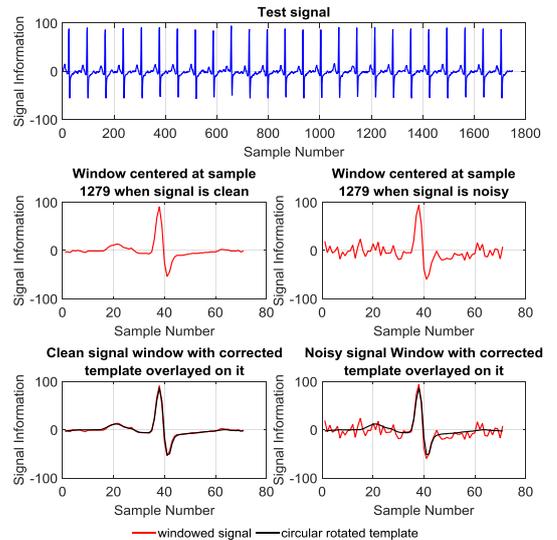

Fig. 3. Figure detailing how for a given signal window centered at a particular point at 2 different SNRs, the template rearranges such that it is adjusted according to the point of maximum correlation and how the cSQI varies depending on noise.

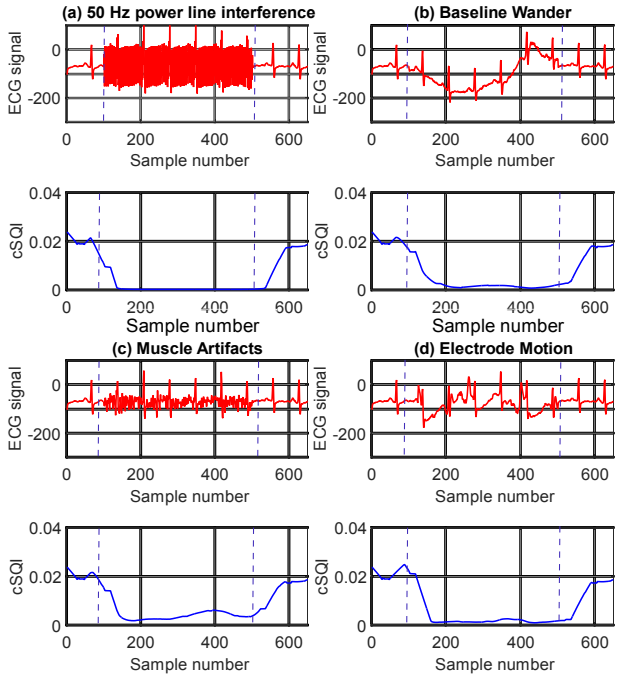

Fig. 4. cSQI plots at each sample point of a signal when it is corrupted by (a) 50 Hz power line interference, (b) Baseline Wander, (c) Muscle Artifacts and (d) Electrode Motion at -10 dB. The vertical dotted lines delimit regions of clean and added noise regions.

The proposed algorithm was also found to have low computational complexity. The complexity of the proposed algorithms is as shown in Tables 3, 4 and 5. From the computational complexity analysis done, it is observable that the smart algorithm discussed is suitable for implementation on a wearable device in real-time.

TABLE 3. Complexity of template generation algorithm (Table 1) for template of length 2M+1 with N cycles

| Operation | Count | Count for N=10, M=35 for $F_s$=125 |
|---|---|---|
| Multiplication | $N(8M^2+6M+1)$ | 100110 |
| Addition | $N(8M^2+4M+1)$ | 99410 |

TABLE 4. Complexity of cSQI calculation during initialization or when calculating for the first c[0]

| Operation | Count | Count for M=35 for $F_s$=125 |
|---|---|---|
| Multiplication | $8M^2+8M+3$ | 10083 |
| Addition | $8M^2+6M+1$ | 10011 |

TABLE 5. Complexity of cSQI calculation for every subsequent sample after initialization

| Operation | Count | Count for M=35 for $F_s$=125 |
|---|---|---|
| Multiplication | $10M+3$ | 353 |
| Addition | $12M+3$ | 423 |

The performance of the cSQI to check for the acceptability of the records in CinC/Physionet challenge 2011 [9] was also carried out. This was done by using the average of the cSQI obtained per channel for a record as features for training a decision tree ensemble. This model used the Adaptive Boosting ensemble creation method with 100 learners for binary classification. For training and validation, Set A of the dataset was used, and the score obtained on validation was 0.938. This model was then tested on the records in Set B of the dataset and a score of 0.904 was obtained in Event 1 of the challenge. These results were found to be not on par with the scores obtained by the top competitors in CinC/Physionet challenge 2011 with the top competitor [19], [20] exhibiting a score of 0.932 in Event 1. But as mentioned in [9], the Physionet challenge used 12 channel ECG data compared to single-channel data used in this work. The top competitors also employed a variety of techniques including machine learning and used a wide range of features to ascertain the acceptability of the records. In this work, we used only one single channel, signal quality indicator-based feature(s) to classify records to aid implementation in an IoT device. This is because this article focuses on the introduction of a novel generalized signal quality indicator for signal quality assessment and the ECG records in the CinC 2011 challenge were used to convey its merit as a robust signal quality indicator on its own. Moreover, the effectiveness of cSQI for instantaneous signal quality indication is observable from Figures 4 and 5.

IV. CONCLUSION

Signal Quality Indicators play an important role in determining whether an acquired signal is useful for further processing and inference. In this article, we propose a novel generalized signal quality indicator that can be used for signal quality assessment of many periodic or quasi-periodic signals provided they have an identifiable fiducial point. This method, after an initial training procedure, is suitable for implementation on a resource constraint wearable device for monitoring biological/physiological signals. In this article, we focus on the suitability of the proposed SQI for quality assessment of ECG signals. It was observed that the proposed SQI meets the requirements for good signal quality indicators, namely, it is a monotonically increasing function of SNR and approximately linear over the SNR region of interest, under all noise scenarios considered. Together with its low complexity make the proposed SQI an attractive option for edge IoT devices.

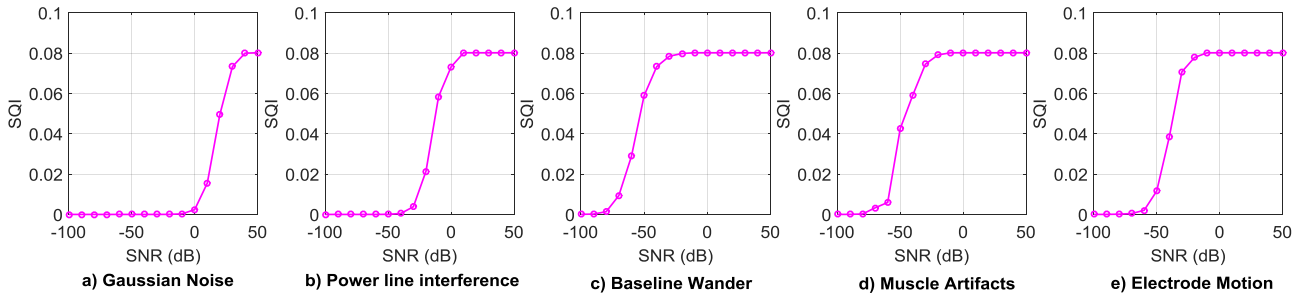

Fig. 5. cSQI vs SNR plot for when the ECG signal is corrupted by (a) Gaussian Noise, (b) 50 Hz power line interference, (c) Baseline wander, (d) Muscle Artifacts and (e) Electrode Motio